\documentstyle[11pt]{article}
\begin{document}
\vspace{0.5 cm}
\begin{flushright}IASSNS 91/76   \\
November 1991
\end{flushright}
\begin{center} {\Large \bf  Once more on the $\theta$-vacua
 in $2+1$ \ \ dimensional \ \  QED and $3+1$ \ \ dimensional gluodynamics.}\\

{\bf A.R.Zhitnitsky}
\footnote{e-mail address is zhitnita@vxcern.cern.ch  }\\
Institute for Advanced Study, Princeton, N.J. 08540,USA. \\
and\\
Institute of Nuclear Physics, Academy of Sciences of the USSR,
630090 Novosibirsk, USSR.
\end{center}
\vspace{1cm}
\begin{abstract}

Two different but tightly connected problems, $U(1)$ and strong
CP violation problems, are discussed in two different models which
exhibit both asymptotic freedom and confinement. One of them is
the 3d Polyakov's model of compact QED and the other is 4d gluodynamics.
It is shown that although both these models possess  the long range
 interactions of the  topological charges, only in the former case
                                                      physics does not
depend on $\theta$; while the latter              exhibits an explicit
$\theta$- dependence.

 The crucial difference is due to the observation, that the
pseudoparticles of 4d gluodynamics possess an aditional quantum
number, apart  of the topological charge        $Q$ .

\end{abstract}

\newpage
{\bf 1.Introduction.}

In the present letter I investigate some problems related to the
$\theta$ vacua in the $SU(2)$ gauge theories. As is known one can add
to the standard Lagrangian the so called $\theta$ term:
\begin{equation}
\label{1}
\Delta L =\theta Q
\end{equation}
where $Q$ is the topological charge which can be written for
the 4d gluodynamics and for 2+1 QED correspondingly as follows:
\begin{equation}
\label{2}
Q=\frac{1}{32{\pi}^2}\int d^4x G_{\mu\nu}^a \tilde{G_{\mu\nu}^a}=
\frac{1}{32{\pi}^2}\int d^4x \partial_{\mu}K_{\mu} , \ \ a=1,2,3.\ \
 \mu , \nu =1,2,3,4.
\end{equation}
\begin{equation}
\label{3}
Q=\frac{1}{8\pi}\int d^3x \epsilon_{ijk}\partial_{i}
(G_{jk}^a {\phi}^a)=\frac{1}{4\pi}\int B_{k}dS_{k} ,\ \ i,j,k=1,2,3.
\end{equation}
Here $G_{\mu\nu}^a$ is the field strength tensor and $\phi^a$ is the scalar
field which is present in the definition of the instanton solution
in the
 2+1 theory (a monopole, in the terminology of the 3+1 theories  )
\cite{Pol1},\cite{Hoof}. In the unitary gauge
$\phi^a$ becomes constant $\phi^a \rightarrow (0,0,1)$ at large
distances  and the combination
$\epsilon_{ijk}G_{jk}^a\phi^a$ reduces to
$B_{i}$. This is just the reason why     this theory is called
the 2+1 QED \cite{Pol2}.

As is known , the $\theta$ term preserves renormalizability of the theory
but is P and T odd. As it can be seen from (\ref{2},\ref{3}) the $\theta$
term is a full divergence, therefore it is equivalent to certain
boundary conditions . The standard line of reasoning
in this case is as follows. Because the boundary conditions can influence
to the physics far from the boundary in the ordered phase only, and
because confinement occurs in a disordered phase , one can think that the all
$\theta$ vacua are equivalent. If  so, the nonvanishing surface integral
(\ref{2},\ref{3}) due to a
 separate topological solution could be an artifact
of the dilute gas approximation which ignores completely the most profound
features of the considering models - confinement. Indeed the surface
integral does not vanish since the topological field falls off too
slowly at large distances              due to the presence of massless
particles in the original Lagrangian , which is not present, however,
in the physical spectrum. Therefore one might hope that the confinement effects
would make the surface integral vanishes. Such a viewpoint based on the
analysis 2+1 QED \cite{Verg}(see also the recent
paper \cite{Samu}on this topic), was strongly advocated by Polyakov.

Indeed , Vergeles has shown \cite{Verg} that as soon as quasiparticles
interaction becomes strong enough, the $\theta$ dependence disappear
from the effective long distance Lagrangian and physics becomes
$\theta$ independent. We will shortly reproduce this result in the next
section ,and emphasize generality of this result and its independence
on
the space dimensions.

At the same time , as is known \cite{Wit},\cite{Ven} ,\cite{Shif}
the $\theta$-
dependence of physics is linked to the $U(1)$ problem.   If we believe
that the resolution of the $U(1)$ problem appears within the framework
of the papers \cite{Wit},\cite{Ven} ,we must assume that the
correlator
\begin{equation}
\label{4}
K=i\int d^4x<0|T{\frac{1}{32{\pi}^2} G_{\mu\nu}^a \tilde{G_{\mu\nu}^a}(x)
,\frac{1}{32{\pi}^2} G_{\mu\nu}^a \tilde{G_{\mu\nu}^a}(0)}|0>
\end{equation}
is nonzero. It means that the vacuum expectation value (vev) of the topological
density
\begin{equation}
\label{5}
<| \frac{1}{32{\pi}^2} G_{\mu\nu}^a \tilde{G_{\mu\nu}^a}|> =
\frac{1}{2}K\theta ,\ \ \ \ \theta \ll 1.
\end{equation}
is nonzero too. But as was shown in ref.\cite{Shif} the nonzero
vev $<G\tilde{G}> \neq 0$ does imply the CP -violation in physical
transition and leads , for example ,to the mixing of the heavy
quarkonium levels with $J^P=0^+$ and $J^P=0^-$ in terms of $<G\tilde{G}>$.
Only dispersion relations are used to translate  $<G\tilde{G}> \neq 0$
into a proof of CP- violation in physical effects.

So, if we believe that $U(1)$ problem is solved in the framework
of Witten-Veneziano approach \cite{Wit},\cite{Ven}
we automatically get $K\neq 0$ (\ref{4}) and therefore,
the nontrivial $\theta$ dependence.

With introduction of light quarks (u,d,s) the value of correlator
$K$(\ref{4})      changes \footnote{It is obviously that $K\sim m_q^1$
irrespectively to the number of light flavours \cite{Ven}.},but
 $\theta$ is still experimentally observable quantity. Moreover,
a few     strict results ( computation of $\eta \rightarrow \pi\pi$ decay
\cite{Shif} and electrical dipole moment \cite{Crew}) were   obtained
using    the soft meson technique.

I will not consider in this papers the theory of light quarks and
confine myself by consideration of pure gluodynamics. I will
discuss the resolution of the aforementioned apparent paradox
(the coexistence of the strong quasiparticle interaction with the
                nontrivial $\theta$ dependence) within framework
of the dynamical toron approach
which was discussed early in context
of different field theories, see ref.\cite{Zhi1} and references therein.
I would like to recall that in all known
 cases, the
toron calculations give, at least, selfconsistent results. Thus, one
may expect that apparent puzzle should be solved
in an automatic way.

 We discuss the statistical ensemble of quasiparticles
which, presumably
  \cite{Zhit}, describes the grand partition function of the
4d YM theory and which possesses long distances strong interaction.
It will be shown that this ensemble describes the system with nontrivial
$\theta$- dependence irrespectively to the strength of  quasiparticle
interaction . Crucial point, in compare with analogous calculation
of ref.\cite{Verg} in Polyakov's model, is a very nontrivial algebraic
structure of the quasiparticle interaction.

Let us now recall some basic facts
           about the toron's role in two and four dimensional physics.
Besides that we    give some arguments why quasiparticles with
 fractional topological charge should be considered in those cases
 and what physical effects arise due to fluctuations with fractional
$Q$.

 The self dual "toron" solution with fractional topological
charge was first considered by 't Hooft\cite{Hoo1} (but in quite
different context). It is                            defined in a
box of size $L_{\mu}$, smeared over this box and exists when the
sizes $L_{\mu}$ satisfy certain relations. The calculation of gluino
condensate $<{\lambda}^2>$ in the supersymmetric YM theory (SYM)
based on the 't Hooft solution was carried out in ref.\cite{Coh}.
However by many reasons (in particular, difficulties with
introduction of fields in the fundamental representation
and  consideration of the ensemble interacting quasiparticles)
the 't Hooft solution \cite {Hoo1} and corresponding calculation
\cite{Coh},\cite{Gom} can be considered as only illustrative example
with fractional topological charge.

Nevertheless,I believe that solutions with a fractional $Q$
may play an important role in theory, but these solutions
should be formulated in another way
(see\cite{Zhi1} and references therein).The corresponding
pseudoparticle
 can be understood as a point defect when the regularization
parameter( which is present in the definition of the solution) goes to
zero. We keep the term 'toron' introduced in ref.\cite{Hoo1}.
By this       we emphasize the fact that the new solution also
 minimizes
the action and carries the topological charge $Q=1/2$,so
it possesses all the characterestics ascribed to the standard
toron\cite{Hoo1}.

All calculations, based on the solution \cite{Zhi1} demonstrate
its  very nontrivial role
for different field theories.
Most glaringly these effects appear in supersymmetric variants of a theory.
In particular, in the
 supersymmetric  $CP^{N-1}$-theories, the torons (point
defects) can ensure a nonvanishing value for the
$<\bar{\psi}{\psi}>\sim\exp(2i\pi k/N+i\theta/N)$ with right
$\theta$-dependence. Such behavior is in agreement with the value
of the Witten index which equals $N$\cite{Wit1} and in agreement
with the large $N$-expansion \cite{Wit2}. In analogous way, the chiral
condensates can be obtained for 4d theories: supersymmetric YM (SYM),
\cite{Coh},
supersymmetric QCD (SQCD). In these cases a lot of various results
are known from  independent consideration (such as the
 dependence of condensates
on parameters $ m,g$; the Konishi anomaly equation and so on...).
\cite{Amat}.  Toron
approach is in agreement with these general results.
The same approach can be used for physically interesting theory of QCD
with $N_f=N_c$. In this case an analogous calculation of $<\bar{\psi}\psi>$
does possible because of cancellation of nonzero modes, like in
supersymmetric theories. For this theory the contribution of the toron
configurations to the chiral condensate has been calculated and is
equal to: $<\bar{\psi}\psi>=-\pi^2\exp(5/12)2^4{\Lambda}^3$ \cite{Zhi1}
(see also E. Cohen \cite{Coh1}).
As is well known in any consistent mechanism for chiral breaking a lot
of problems, such as: the $U(1)$-problem, the number of discrete vacuum
states, the $\theta$-puzzle, low energy theorems and so on, must be solved
in an automatic way. We have checked that
 all these properties \cite{Zhi1}  are
                             consistent with    the toron calculation.

To give some insight about the
       algebraic structure of the quasiparticle interaction, I
would like to cite some results for $2d CP^{N-1}$ model which is known
\cite{Wit2} possesses nontrivial $\theta$-dependence. It turns out
that the toron gas contribution to grand partition function
in $ 2d CP^{N-1}$ model reduces to the classical Coulomb system(CCS)
 \cite{Zhit}:
\begin{eqnarray}
\label{6}
Z=\sum_{k=0}^{\infty} \frac{{\lambda}^{k_1+k_2}}{(k_1)!(k_2)!}
\sum_{{\mu_{\alpha}},{q_{\alpha}}}\prod_{i=1}^{k_1+k_2}d^2x_i
exp(-\epsilon_{int.}) ,~~~~~~~~~~~~~~~~~~~~~~~~~        \\
\epsilon_{int.}=-4\sum_{i>j}q_i \vec{\mu_{i}} ln(x_i-x_j)^2
q_j \vec{\mu_{j}} +2\ln L^2(\sum_{i}q_i\vec{\mu_i})^2, \ \ \ \
\lambda =c \frac{M_0}{f(M_0)} exp(-\frac{\pi}{f(M_0)}) .  \nonumber
\end{eqnarray}
where N different kinds of torons classified by the weight
$\vec{\mu_{\alpha}}$ of fundamental representation of the
$SU(N)$ group and $q_i$ is the sign of the topological charge.
Besides that, in formula (\ref{6}) the value $f(M_0)$ is the bare
coupling constant and $M_0$ is ultraviolet regularization, so that
in eq.(\ref{6}) there appears the renormalization invariant combination
$\lambda$. As can be seen from (\ref{6}) the configurations
only satisfying the neutrality requirement
\begin{equation}
\label{6a}
\sum_i q_i \vec{\mu_i}=0
\end{equation}
are essential in thermodynamic limit $L \rightarrow \infty$.
However we will consider the system in the box with size $ L$  and so we keep
this term for the future analysis.

Using the correspondence between CCS and the Toda field theory
\begin{eqnarray}
\label{7}
Z_{\theta}=\int D\vec{\phi} exp(-\int d^2x L_{eff.}) ,~~~~~~~~~~~~~~~~~~~~~~~~
{}~~~~~~~~~~~~~~~~~~~~~~~~~~~~~~~\\
L_{eff}=1/2(\partial_{\mu}\vec{ \phi})^2-\sum_{\vec{\mu_{\alpha}}}
{\lambda}^2\exp(i4\sqrt{\pi}\vec{\mu_{\alpha}}\vec{\phi}+i\theta/N)
-\sum_{\vec{\mu_{\alpha}}}{\lambda}^2\exp
(-i4\sqrt{\pi}\vec{\mu_{\alpha}}\vec{\phi}-i\theta/N). \nonumber
\end{eqnarray}
the expectations of different values( the vacuum energy, the
topological
density, the Wilson line ...) was calculated. All results ( confinement,
right dependence on $\theta/N$ and so on ) are precisely what one obtains
 from the large N expansion.
In this effective field theory ,$\vec{\phi}$ represents the
$N-1$ component scalar potential , the sum over $\vec{\mu_{\alpha}}$
in (\ref{7}) is over the $N$ weights of the fundamental representation of
$SU(N)$ group. Note, that the first interaction term is related to
torons and the second one to
     antitorons. Besides that, since we wish to
 discuss the $\theta$ dependence , we also include a term proportional
 to the topological charge density $\frac{\theta}{4\pi}\epsilon_{\mu\nu}
F_{\mu\nu}$ to the starting lagrangian (formula, analogous to eqs.(\ref{1},
\ref{2},\ref{3})), and corresponding track from this to the effective
lagrangian(\ref{7}).

The
most important result from ref.\cite{Zhit}
is the nontrivial dependence on $\theta$ of the topological density
and susceptibility, the values which are relevant for the solution
of the $U(1)$ problem :
\begin{equation}
\label{8}
< \frac{\epsilon_{\mu\nu}F_{\mu\nu}}{4\pi}>_{\theta}
\sim \sin(\frac{\theta}{N}).
\end{equation}
\begin{equation}
\label{9}
\int d^2x< \frac{\epsilon_{\mu\nu}F_{\mu\nu}}{4\pi}(x),
 \frac{\epsilon_{\mu\nu}F_{\mu\nu}}{4\pi}(0)>
\sim \frac{1}{N} \cos(\frac{\theta}{N}).
\end{equation}
Here $\frac{F_{\mu\nu}\epsilon_{\mu\nu}}{4\pi}$ is the topological density
in $2d CP^{N-1}$ model. The vacuum characteristics listed above
are analogous to eqs.(\ref{4},\ref{5}) in 4d YM theory.

The reason to have the nontrivial $\theta$- dependence (\ref{8},\ref{9})
as well as the strong quasiparticle interaction $\sim \ln(x_i-x_j)^2$
(\ref{6}) in this $2d CP^{N-1}$ model is the presence of the nontrivial
algebraic structure $\sim q_i \vec{\mu_{i}} q_j \vec{\mu_{j}} $
in the expression for $\epsilon_{int}$ (\ref{6}). Just this fact
was crucial in the analysis of the $2d CP^{N-1}$ model. As we observed early,
\cite{Zhit} the same structure for quasipartcle interaction take place
in the 4d YM theory in the striking contrast with 2+1 Polyakov's model
\cite{Pol2}, where the interaction energy proportional to the topological
(magnetic) charges (\ref{3}) of quasiparticles only:
\begin{equation}
\label{10}
\epsilon_{int}\sim \frac{q_i q_j}{|x_i-x_j|}.
\end{equation}
Just this difference leads, as will be shown bellow, to the existence
of the nontrivial $\theta$ dependence in 4d YM theory in spite of the
fact of the strong quasiparticle interaction $\sim \ln(x_i-x_j)^2$.

{\bf 2.Comparative analysis of the 2+1d QED and 3+1d gluodynamics}.

We start from the short review of the results of ref.\cite{Verg}
to demonstrate the independence of $\theta$ the partition function in
Polyakov's model \cite{Pol2}            . To do this let us consider
the contribution to the partition function of the configurations
 with fixed
numbers of $n_1$ monopoles and $n_2$ antimonopoles. As has been shown
by Polyakov \cite{Pol2} the corresponding contribution reduces
to the following expression :
\begin{eqnarray}
\label{11}
Z_{\theta}^{NQ}\equiv \exp(iQ\theta)\exp(-F_{NQ})=\exp(iQ\theta)
\frac{1}{n_1! n_2!}\prod_{i=1}^{N}d^3x_i\lambda^N\exp(-
\sum_{i\neq j}^N\frac{q_iq_j}{|x_i-x_j|})  \\   \nonumber
Q=n_1-n_2,\ \ \ N=n_1+n_2, \ \ \ q_i=\pm 1,\ \ \ i=1,N
\end{eqnarray}
Here ,           $\lambda$ is fugacity of the obtained Coulomb gas
and it is
 determined by the ultraviolet behaviour of the theory.

                              If the Coulomb interaction could be ignored,
then (\ref{11})       yields for free energy $F_{\theta}$ the $\theta$-
dependent expression. This is a well known result obtaining in the framework
of dilute noninteracting gas approximation.

In our case the partition function with a fixed topological charge
\begin{equation}
\label{12}
Z_{\theta}^Q\equiv \exp(iQ\theta)\exp(-F_Q)=\sum_{N}Z_{\theta}^{NQ}
\rightarrow \int dN Z_{\theta}^{NQ}
\end{equation}
can be obtained by the steepest descent method with respect $N$.
With $L$ ,the linear size of the system, going to infinity, we get
the following expression for the noninteracting case:
\begin{eqnarray}
\label{13}
\exp(-F_Q)=\int dN \exp[N\ \ ln(L^3\lambda)-\frac{N+Q}{2}(ln\frac{N+Q}{2}-1)
-\frac{N-Q}{2}(ln\frac{N-Q}{2}-1)]\sim  \\    \nonumber
\exp(N_0)\exp(-\frac{Q^2}{N_0}), \ \ \ \ \ \ \ \ \ \ \ \ \ \ \ \ \ \ \ \ \
\ \ \ \ \ \ \ \ \ \ \ \ \ \ \ \ \ \ \ \ \ \
\end{eqnarray}
where we have used formula $\ln(k!)\sim k\ln k-k$ and
  $N_0$ is to be determined from the equation :
\begin{equation}
\label{14}
ln(L^3\lambda)=\frac{1}{2}ln(\frac{N_0^2-Q^2}{4}), \ \ \ \
\frac{Q}{N}\ll 1, \ \ \ \ N_0 \simeq2L^3\lambda .
\end{equation}
Now the partition function for the dilute noninteracting
gas is evaluated in the following way:
\begin{eqnarray}
\label{15}
Z_{\theta}\equiv\exp(-F_{\theta})\sim\exp(2L^3\lambda)\int dQ
\exp(i\theta Q)\exp(-\frac{Q^2}{4\lambda L^3}) \ \ \sim
\exp[2\lambda L^3(1-\frac{\theta^2}{2})] \\    \nonumber
F_{\theta}\simeq -2\lambda L^3 (1-\frac{\theta^2}{2}), \ \ \ \ \ \ \ \ \
\theta \ll 1   \ \ \ \ \ \ \ \ \ \ \ \ \ \ \ \ \ \ \ \ \ \ \ \ \ \ \ \ \
\end{eqnarray}
As it can be seen from(\ref{15}) the essential terms are those
which have the charges
\begin{equation}
\label{15a}
|Q|\sim\lambda L^3\theta, \ \ \ \ \ \ \ \ \ \frac{|Q|}{N}\sim\theta
\end{equation}
and our approximation $Q/N \rightarrow 0$ can be justified in the
limit $\theta \rightarrow 0$.
So, as        expected , the free energy $F_{\theta}$ does depend
on $\theta$ explicitly.

Now we return to the Polyakov's model. It is known that the
Coulomb interaction in plasma cannot be ignored. Hence, formula
(\ref{15}) is not true. To give some estimation in this case
let us place our system to the box with size $L$. In this case,
the excessive charge deposits on the wall of the box and the
free energy is the same as for neutral Debye
plasma plus the Coulomb energy \cite{Verg}:
\begin{equation}
\label{16}
\exp(-F_Q)\sim\int dN \exp[N\ \ ln(L^3\lambda)-\frac{N+Q}{2}(ln\frac{N+Q}{2}-1)
-\frac{N-Q}{2}(ln\frac{N-Q}{2}-1)-\frac{Q^2}{L}]
\end{equation}
Now the partition function with a fixed charge can be obtained by the same way
as before (\ref{13}).          The estimates  of $F_{\theta}$ is now  less
trivial:
\begin{equation}
\label{17}
Z_{\theta}\equiv\exp(-F_{\theta})\sim\exp(2\ \ L^3\lambda)\int dQ
\exp(i\theta Q)\exp(-\frac{Q^2}{4\lambda L^3})\exp(-\frac{Q^2}{L})
\end{equation}
Here the new factor $\exp(-\frac{Q^2}{L})$  is
due to the Coulomb interaction,and now
     the essential configurations                      have charges:
\begin{equation}
\label{18}
|Q|\sim L\theta ,\ \ \ \ \ \ \frac{Q}{N}\sim\frac{\theta}{\lambda L^2}
\rightarrow 0
\end{equation}
and therefore                               :
\begin{equation}
\label{19}
F_{\theta}\sim -2\lambda L^3+L\theta^2=-2\lambda L^3(1-\frac{\theta^2}
{2\lambda L^2}).
\end{equation}
Thus, at $L\rightarrow \infty$ the free energy $F_{\theta}$
in the Polyakov's model does not depend on $\theta$ and coincides
with the corresponding expression at $\theta=0$. Let me   emphasize
that                            this result is
due to the strong interaction of quasiparticles.

At first sight this derivation looks rather general.
Moreover, one can suspect that as soon as we have a strong enough
interaction and corresponding trace of it in the formula analogous to
(\ref{17}),we will obtain the expression  for the free energy,  like
(\ref{19}),which does not exhibit any $\theta$ dependence at
$L\rightarrow\infty$. This is indeed the case for the simplest
algebraic structure $\sim q_iq_j$ for  the quasiparticle interaction.
Moreover, the arguments, given above are in perfect agreement
with intuitive picture ( discussed in Introduction) connecting
$\theta$ independence of the effective lagrangian and    confinement.

Let us now proceed  to the analysis of 4d gluodynamics.
In this case, like in $ 2d CP^1$ model, a toron classified by two
numbers :    the sign of the topological charge $q_i$ and    the
isotopic   projection $I_i$:
\begin{equation}
\label{1a}
|q_i,I_i>
\end{equation}
The details are in the original paper \cite{Zhit} ,             and
     I would like only    mention here    the definition of the toron
isotopic projection.

Let us introduce along with papers \cite{Loos} an additional
object into the theory, so called measuring operator $C$
such that
\begin{equation}
\label{2a}
D_{\mu}C(x)=0
\end{equation}
and consider the integral
\begin{equation}
\label{3a}
I=\int tr(CF_{\mu\nu})d\sigma_{\mu\nu}
\end{equation}
where the toron lies in a plane $\sigma_{\mu\nu}$.
In this case the different choice of $C$ obeing eq.(\ref{2a})
yelds different isospin directions. It is useful to keep in mind
some anology with monopole's classification. In this case
the role of $C$ plays      a Higgs field \cite{Oliv},
which far from the core satisfies the same equation (\ref{2a}).
Let us note, that an each quasiparticle is classified by
$I_3$ proection and so has a    nontrivial transformation properties
under the
 $SU(2)$ gauge group. However due to the neutrality condition in the
infinite volume limit (\ref{6a}), the vacuum is the singlet state
under the gauge transformation. Moreover,the answer does not depend
on the choice of the $z$ axis (the axis of quantization)when the sum
over all possible isospins for all quasiparticles will be done.

With this in mind let us consider  in  more details the grand
partition function for the 4d YM theory \cite{Zhit} which looks like
(\ref{6}) with some trivial changes :
\begin{eqnarray}
\label{20}
\lambda_{CP^{N-1}}\rightarrow \lambda_{YM}\sim
(\frac{M_0^{11/3}}{g^4(M_0)}\exp(-\frac{4\pi^2}{g^2(M_0)}))^{\frac{3}{11}}
\ \ \ \ \ \ \ \ \ \ \ \ \ \ \ \ \ \ \ \  \\  \nonumber
d^2x\rightarrow d^4x ,\ \ \ \ \ \ \ \ \ \ \
-4\ln(x_i-x_j)^2\rightarrow -\frac{2}{3}\ln(x_i-x_j)^2 ,\ \ \ \ \ \
-4\ln L \rightarrow -\frac{2}{3}\ln L .
\end{eqnarray}
The last replacements in          (\ref{20})
                         is the direct consequence of renorminvariance
of the theory \cite{Zhit}. Now let us examine the contribution
to the partition function of the terms with fixed numbers of
$n_1(m_1)$ torons (antitorons) with isospin $up$
and $n_2(m_2)$ torons (antitorons) with isospin $ down$. In this case
the integration over $N$ can be done as before by the steepest
descent method with respect $N$ ( let us note that the term
related with interaction does not depend on $N$) and so the
formula analogous to eq.(\ref{17}) for the free energy $F_{\theta}$
in 4d gluodynamics looks as follows:
\begin{eqnarray}
\label{21}
Z_{\theta}\equiv\exp(-F_{\theta})\sim\exp(2L^4\lambda)\int dQ
\exp(i\frac{\theta}{2}Q)\exp(-\frac{Q^2}{4\lambda L^4})
\int d(Q_1-Q_2)\exp(-\frac{2}{3}(Q_1-Q_2)^2\ln L)  \\   \nonumber
Q\equiv (Q_1+Q_2),\ \ n_1+n_2=1/2(N+Q_1+Q_2),\ \ m_1+m_2=
1/2(N-Q_1-Q_2)
\end{eqnarray}
Here the $\frac{1}{2}Q_1=\frac{1}{2}(n_1-m_1) $ is the topological charge
carried by the quasiparticles with isospins $up$ and
 $\frac{1}{2}Q_2=\frac{1}{2}(n_2-m_2) $
is the same for quasiparticles with isospins $ down$.
While   obtaining                   (\ref{21}) we took into account
the form of the last term in eq.(\ref{6}) (with corresponding
replacement (\ref{20}) for transition from $CP^{N-1}$ model
to 4d gluodynamics):
\begin{equation}
\label{22}
(\sum_i\mu_i q_i)^2\ln L \rightarrow (n_1-m_1-n_2+m_2)^2\ln L=
(Q_1-Q_2)^2 \ln L.
\end{equation}
The trace of this interaction is the appearance of the last term
in eq.(\ref{21}) analogous to the corresponding contribution
$\sim \exp(-\frac{Q^2}{L})$ in eq.(\ref{17}) for 2+1 QED. Besides that,
                                        we have the factor
$(i\frac{\theta}{2})$ in eq.(\ref{21}) instead of$ (i\theta)$ in
eqs.(\ref{15},\ref{17}): it is the direct consequence of
the fractional value for the toron topological charge equals one half.

Now we see the crucial difference between the expressions for free energy
in the 4d gluodynamics (\ref{21}) and 2+1d QED (\ref{17}).
If the topological quasiparticles were classified by topological charge only
(it would corresponds $Q_2=0$ in the formula (\ref{21})) we would
obtain the $\theta$ independent expression for free energy, just as it
has happened in 2+1d QED , see eq.(\ref{19}).

But fortunately , we have less trivial expression for 4d gluodynamics:

         The last term
                 does not depend on combination $ Q_1+Q_2$ and is
                  factorized
out . The nontrivial on $\theta$ integral over $d(Q_1+Q_2)$ is reduced
to the noninteraction gas case (\ref{15}) and does depend on $\theta$
explicitly at $L\rightarrow\infty$:
\begin{equation}
\label{23}
F_{\theta}\simeq -2\lambda L^4 (1-\frac{1}{2}(\frac{\theta}{2})^2), \ \ \ \ \ \
 \ \ \
\theta \ll 1   \ \ \ \ \ \ \ \ \ \ \ \ \ \ \ \ \ \ \ \ \ \ \ \ \ \ \ \ \
\end{equation}
Let me   emphasize that this result   has been  obtained
 due to the
relevant new quantum number classifying the
 quasiparticle.
Besides that I interpret this result as the expansion of
\begin{equation}
\label{4a}
F_{\theta}\sim -|\cos(\frac{\theta}{2})|
\end{equation}
obtained in \cite{Zhit} by quite different method.
As discussed in that paper the reason for such $\theta$
dependence is existence of the two vacuum solutions (for
$SU(2)$ group) minimizing vacuum energy. Althogh each of the
separate solutions has a $\theta$ period of $4\pi$, the overall
minimum has a $\theta$ period of $2\pi$ because the solutions
jump from one value to another at $\theta=\pi$. Very important
  is that two solutions have been prepaired for such jumps from
the very beginning. Indeed, as soon as we allowed one half topological
charge,the number of the classical vacuum states is increased by the same
factor two in compare with a standard classification,
counting only integer winding numbers $|n>$.

Of course,vacuum transitions eliminate this degeneracy.
However the trace of enlargement number of the classical vacuum
states does  not dissapear. Vacuum states now classified by two
numbers : $0\leq \theta <2\pi$ and $k=0,1$. Let me repeat that
origin for this is our main assumption that fractional charge
is admitted and therefore the number of classical vacuum
states is multiplied by a factor two.

I have to note that the same situation takes place in the
supersymmetric YM theory, but in this case the vacuum states
are still degenerate after vacuum transitions. The number $k$
in this case just numerates different vacua at the same
$\theta$. However, the gauge classification for the
winding vacua before perturbation
 for both models  (supersymmetric and nonsupersymmetric one)
is the same.

We close this discussion by  a remark    that the analogous $\theta/N$
dependence was discovered in gluodynamics at large N
\cite{Wit},\cite{Ven},
\cite{Wit3}. In these papers was argued that the vacuum
energy at large N appears in the form $E\sim E(\theta/N)$.
Such a function can be periodic in $\theta$ with period $2\pi$
only if there are many vacuum states for given values of $\theta$.
Indeed, gluodynamics can be understood as a QCD with very large
quark's mass. In this limit effective lagrangian can be founded
\cite{Wit3} and it turns out that the number of vacua
is of order $N$ at $N\rightarrow\infty$. This fact actually
is coded in the effective lagrangian containing the
multi-branched logarithm $\log\det (U)$.
In the Veneziano approach \cite{Ven} the same fact can be seen
from the formula for multiple derivation of  the topological
density $Q$ with respect to $\theta$ at $\theta =0$.
\begin{equation}
\label{a}
\frac{\partial^{2n-1}}{\partial\theta^{2n-1}}<Q(x)>\sim
(\frac{1}{N})^{2n-1} \ \ ,n=1,2...
\end{equation}
 So, eq.(\ref{4a})
definitely does not contradict to results for large
N expansion.

It is instructive to understand this result in terms of the effective
field description  . Just like in $ 2d CP^{N-1}$ model, the
grand partition function for toron gas in 4d gluodynamics can be
re-expressed in terms of effective field theory very similar to eq.
(\ref{6}). The difference with $2d CP^{N-1}$ model only in the
kinetic term, which has the standard form $\sim(\partial_{\mu}\phi)^2$
in $2d CP^{N-1}$ model and looks more complicated in 4d gluodynamics
$\sim (\Box\phi)^2$ \cite{Zhit}. This difference, however, does not
influence on the arguments given bellow. Only the interaction part
of the effective lagrangian , depending on $\theta$, is relevant for
our analysis.

For Polyakov's model the effective lagrangian has the same
Sine-Gordon form (\ref{7}), but with a crucial difference.
Namely , the grand partition function (\ref{11}) for 2+1d QED
can be re-expressed in terms of a field theory with lagrangian
\begin{equation}
\label{24}
L_{eff}=\frac{1}{2}(\partial_{\mu}\phi)^2-\lambda\cos(\phi+\theta)
\end{equation}
but without any additional summing over$\mu_{\alpha}$
\footnote{Let us note, that in 4d gluodynamics this summing  reproduces
in the effective Lagrangian the
 well known $Z_N$ Weyl symmetry  of the original YM theory.}.
After shift $\phi\rightarrow\phi-\theta$, the $\theta$ dependence
in the $L_{eff}$ disappear in according with our formula
(\ref{19}),\cite{Verg}.

In contrast with it
 the situation in the $2d CP^{N-1}$ model and 4d gluodynamics is quite
different because in these cases the $\theta$ parameter can not be removed
from effective lagrangian (\ref{7}) in agreement with formula (\ref{23}).

{\bf 3.Final remarks}.

The main point of this Letter is that the $ dynamical$ resolution of
the $U(1)$ problem in gluodynamics looks very naturally in
terms of ensemble of the quasiparticles
with the fractional  topological charges (torons),which have a
         nontrivial
long range interactions                 . This approach is in a perfect
agreement with Witten-Veneziano solution of the $U(1)$ problem
 in framework of large N-expansion
\cite{Wit},\cite{Ven}. Besides that this approach demonstrates
a self-consistency of the different calculation in a various
field theories \cite{Zhi1},where the results are known beforehand.

We have discovered that the algebraic structure of the
quasiparticle interaction  is just what is needed for the
solution of $U(1)$ problem irrespectively to the strength of interaction.
This result in gluodynamics is in a striking     contrast with
that for the 2+1 QED, where strong quasiparticle interaction
eliminates the $\theta$ dependence in the theory.

The other point which I would like to mention here is as follows.
Although we have discussed the dynamical solution of the $U(1)$ problem
within the toron framework          , the obtaining results have
    a more general origin. Indeed ,  we have shown
that the nontrivial $\theta$ dependence $ does\ \ appear $
in the effective lagrangian \footnote{This is an essential
requirement for the solution of the $U(1)$ problem} in spite of the
fact of strong quasiparticle interaction, $provided \ \ the\ \
 quasiparticles \ \ are \\
\ \ classified\ \ by\ \  the\ \ new\ \ quantum
\ \ number$                           ,the weight of representation of
the group.

{\bf Acknowledgments}

I am very grateful to Frank Wilczek for stimulating
discussions and for hospitality at IAS, Princeton,
where this manuscript was finished. I thank I.B.Khriplovich for
several instructive discussions during the course of this work
and S.Samuel for sending me his recent preprint\cite{Samu}
{}.

\end{document}